\documentclass[aps,pra,twocolumn,superscriptaddress]{revtex4-2}
\usepackage{mathtools}
\usepackage{amsmath,amscd,amssymb,amsthm}

\usepackage{xcolor}
\begin{document}

\title{Enhanced-sensitivity interferometry with phase-sensitive unbiased multiports}

\author{Christopher R. Schwarze}
\email[e-mail: ]{{\tt crs2@bu.edu}}
\affiliation{Department of Electrical and Computer Engineering \& Photonics Center, Boston University, 8 Saint Mary’s St., Boston, Massachusetts 02215, USA}
\author{David S. Simon}
\email[e-mail: ]{{\tt simond@bu.edu}}
\affiliation{Department of Electrical and Computer Engineering \& Photonics Center, Boston University, 8 Saint Mary’s St., Boston, Massachusetts 02215, USA}
\affiliation{Department of Physics and Astronomy, Stonehill College, 320 Washington Street, Easton, Massachusetts 02357, USA}
\author{Alexander V. Sergienko}
\email[e-mail: ]{{\tt alexserg@bu.edu}}
\affiliation{Department of Electrical and Computer Engineering \& Photonics Center, Boston University, 8 Saint Mary’s St., Boston, Massachusetts 02215, USA}
\affiliation{Department of Physics, Boston University, 590 Commonwealth Avenue, Boston, Massachusetts 02215, USA}


\begin{abstract}
Here we introduce interferometric devices by combining optical feedback (cavities) with unbiased multiports, which unlike traditional beam dividers, allow light to reflect back out of the port from which it originated. By replacing the traditional, directionally-biased beam-splitter in a Michelson interferometer with an unbiased multiport, {\color{black} the functional dependence of the scattering amplitudes changes.} As a result, the derivative of transmittance with respect to an external phase perturbation can be made substantially large. This significantly enhances the resolution of phase measurement{\color{black}, and allows the phase response curves to be altered in real time by tuning an externally-controllable phase shift.}
\end{abstract}
\maketitle

\section{Introduction\label{intro}}
Beam dividing is a fundamental manipulation of light, forming the basis of applications such as interferometry \cite{Born:382152} and optical information processing \cite{RevModPhys.79.135}. {\color{black} Many optical apparatuses rely on a variant of the standard beam-splitter to separate and combine light. This might be a cube beam-splitter, a pellicle one, or perhaps a fiber coupler. In any case, these devices share the following characteristic: the light entering a given port cannot exit that same port. This directionally-biased behavior effectively lowers the dimensionality of the device: an output state is unable to populate as many modes as there are ports due to the lack of a coupling between the input port and the mode counter-propagating from that port. Sometimes this feed-forward nature is desirable, for instance, to prevent back-reflections from re-entering a laser cavity. However, the use of strictly feed-forward scattering devices constrains the output states that can be generated, and consequently, limits the capabilities of the systems that use them.}

A generalization to the class of feed-forward linear-optical scatterers was introduced in \cite{PhysRevA.93.043845}. These new devices are directionally-\textit{unbiased}, allowing light to reflect out the port that it entered. In addition to reducing the number of optical elements needed to enact certain operations, this property has driven a number of theoretical developments in entangled-state information processing \cite{PhysRevA.102.063712}, Hamiltonian simulation of topologically nontrivial systems \cite{Feldman_2007}\cite{PhysRevA.95.042109}, and optical sensing with the Sagnac effect\cite{PhysRevA.106.033706}.

{\color{black} In this article, we use a particular directionally-unbiased counterpart of the beam-splitter called the Grover ``$N$-sided coin'', which will be introduced in greater detail later. With the four-port Grover coin, we form a generalization of the Michelson interferometer which can be configured to have an arbitrarily large sensitivity to a phase perturbation. This behavior stems from the optical cavities which get formed by replacing the beam-splitter with the unbiased Grover coin. Intentionally leaky cavities of this sort formed the basis of the free-space realization of the 3-port Grover coin \cite{Osawa:18}. In this case, we analyze how a controllable phase shift acquired during each round trip in the cavity can be used to create a lower-dimensional device with a tunable scattering matrix.}

The outline for the remainder of the article is as follows. In the next section, we review linear-optical scattering theory for monochromatic radiation interacting with devices such as the beam-splitter or Grover coin. In Section III, we compare various unbiased and tunable scattering devices formed from either a single beamsplitter or a 4-port Grover coin. All devices considered are tuned by changing optical phase elements; that is, for new values of each phase shift, a new scattering matrix is produced which is periodic in $2\pi$. In Section IV, we focus on a two-cavity configuration which resembles the Michelson interferometer. Unlike the standard Michelson interferometer, this Grover-based version can be tuned to have a transmittance with an arbitrarily steep slope, which can be used to obtain enhanced sensing. Conclusions are drawn in Section V.

\section{Linear-optical scattering theory}

We will mathematically represent a beam-splitter as a particular linear, spatially-coherent scattering transformation. Any optical device that can be expressed this way we will call a ``multiport''. The action of a multiport is often expressed in the ideal case as a unitary transformation, which operates on the probability amplitudes of an optical state. It is common to express such a state in a Fock basis of spatial modes. 

\begin{figure}[ht]
\centering\includegraphics[width=0.45\textwidth]{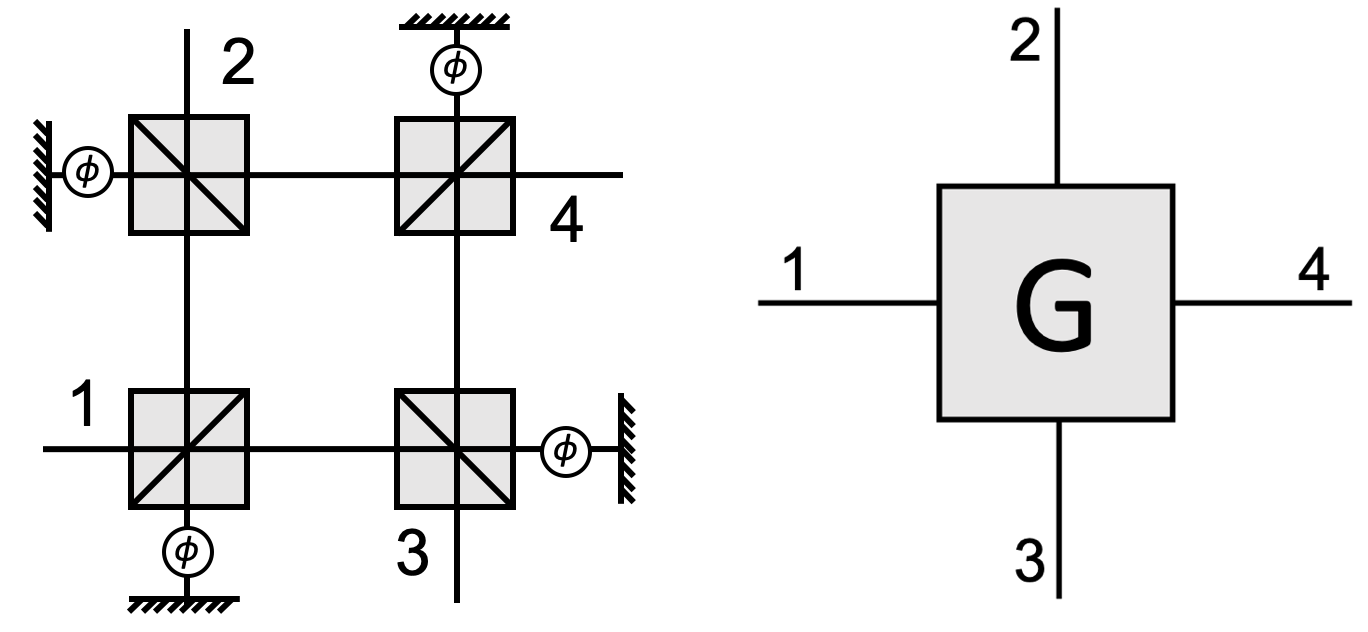}
\caption{(left) One particular realization of a directionally-unbiased optical multiport, formed by linking four 50:50 beam-splitters to form a leaky cavity. The distance between beam-splitters is twice that between a mirror and neighboring beam-splitter. By fixing $\phi = -3\pi/4$, so that the total phase per mirror-unit is $-\pi/2$, one obtains the four-port Grover coin \cite{PhysRevA.95.042109}. We abstract this four-port device into the circuit diagram (right), which will be used throughout the rest of the article. 
\label{fig:grover-real}}
\end{figure}

If we consider this basis and additionally assume all excitations of interest are confined to four {\color{black} ports}, a general superposition of arbitrarily polarized, monochromatic radiation takes the form {\color{black} $|\psi\rangle =  (c_1a_1^\dagger + c_2a_2^\dagger + c_3a_3^\dagger + c_4 a_4^\dagger )|0\rangle$. The subscripts on the creation operators indicate the port numbers of a given four-port device, such as those depicted in Fig. \ref{fig:grover-real}. The field modes could be fiber modes or free space plane waves, depending on the physical implementation of the system.} Next we identify each {\color{black} photon creation} operator $a_j^\dagger$ with the standard basis vector $e_j$, which equals $1$ at its $j$th element and $0$ elsewhere. {\color{black} In our notation, ingoing and outgoing modes of the same port will be associated with the same standard basis vector.}

{\color{black} In this formalism}, the state $|\psi\rangle$ may be expressed as a column vector
\begin{equation}
    |\psi\rangle = 
    \begin{pmatrix}
     c_1\\
     c_2\\
     c_3\\
     c_4
    \end{pmatrix}.
\end{equation}
while a 50:50 beam-splitter would have a scattering matrix given by
\begin{equation}\label{eq:bs}
B = \frac{1}{\sqrt{2}}
\begin{pmatrix}
0 & 0 & 1 & 1\\
0 & 0 & 1 & -1\\
1 & 1 & 0 & 0\\
1 & -1 & 0 & 0
\end{pmatrix},
\end{equation}
and the output state of $|\psi\rangle$ interacting with the beam-splitter is then given by $B|\psi\rangle$\ \cite{BS1}\cite{PRASAD1987139}. 


The transformation (\ref{eq:bs}), however common in practice, is relatively sparse: modes 1 and 2 are only routed to modes 3 and 4 and vice versa; no energy returns to the mode from which it came. In this sense, the device is feed-forward or directionally-biased. Hence the scattering matrix above is often depicted by the $2 \times 2$ matrix
\begin{equation}
H = \frac{1}{\sqrt{2}}
\begin{pmatrix}
 1 & 1\\
 1 & -1
\end{pmatrix}
\end{equation}
and the feed-forward nature is assumed implicitly. 

When static scattering devices such as the beam-splitter are combined with optical phase shift elements, the resulting device may be viewed as a realization of a tunable scattering matrix. Non-trivial device behavior is often obtained with interference, such as when excitations in multiple spatial modes are superimposed after a tunable delay is introduced between them. To obtain exotic forms of interference, an increasing number of optical elements and/or degrees of freedom are typically required. However, as we will discuss later, interference across an infinite number of cavity paths can give rise to novel effects without substantially increasing either.


An important instance of an unbiased multiport is the Grover coin, which originates from a matrix that appeared in Grover's search algorithm \cite{grover} and has since been adopted to studies of quantum walks \cite{qws}. The $d$-port Grover coin (where $d \geq 3$) is defined to have a reflection amplitude of $(2/d - 1)$ and transmission amplitude of $2/d$ at each output port, for all input ports. Therefore in addition to being unitary, it is also real and symmetric. Grover coins have been experimentally realized for the case of $d = 3$ \cite{Osawa:18}\cite{Kim:21} and fabrication of an integrated version of the $d = 4$ case is underway. 

In the $d = 4$ case, the scattering matrix $G$ is given by
\begin{equation}\label{eq:grover}
    G = \frac12
    \begin{pmatrix}
    -1 & 1 & 1 & 1\\
    1 & -1 & 1 & 1\\
    1 & 1 & -1 & 1\\
    1 & 1 & 1 & -1\\
    \end{pmatrix}.
\end{equation}

As in Eq. (\ref{eq:bs}), the output probabilities are equal at each port. {\color{black} The free-space realization of the device, first presented in \cite{PhysRevA.95.042109}, is shown next to its abstract circuit symbol in Fig. \ref{fig:grover-real}. {\color{black} In order for this realization to possess the scattering matrix (\ref{eq:grover}), the phase imparted in each arm $\phi$ must equal $-\pi/2$. Moreover, the source of radiation must be coherent with respect to the length scales of the beam-splitter arrangement. This prevents various internal routes that a photon could propagate along from being distinguished by the photon exit times, causing the amplitudes corresponding to these paths to interfere. These coherence relations are further discussed in \cite{PhysRevA.93.043845}. }

The permutation-symmetry of the matrix (\ref{eq:grover}) manifests as a rotational symmetry in the device in Fig. \ref{fig:grover-real} (left).} Since this Grover coin and the beam-splitter both have four ports, it is interesting to consider what results when traditional beam-splitters are replaced with Grover 4-ports in conventional interferometers. It has recently been shown that in the Mach-Zehnder topology, the Grover coin enables simultaneous measurements of different phase shifts \cite{PhysRevA.106.033706}. 

\section{Grover Optical Cavities}

\begin{figure}[ht]
\centering\includegraphics[width=.45\textwidth]{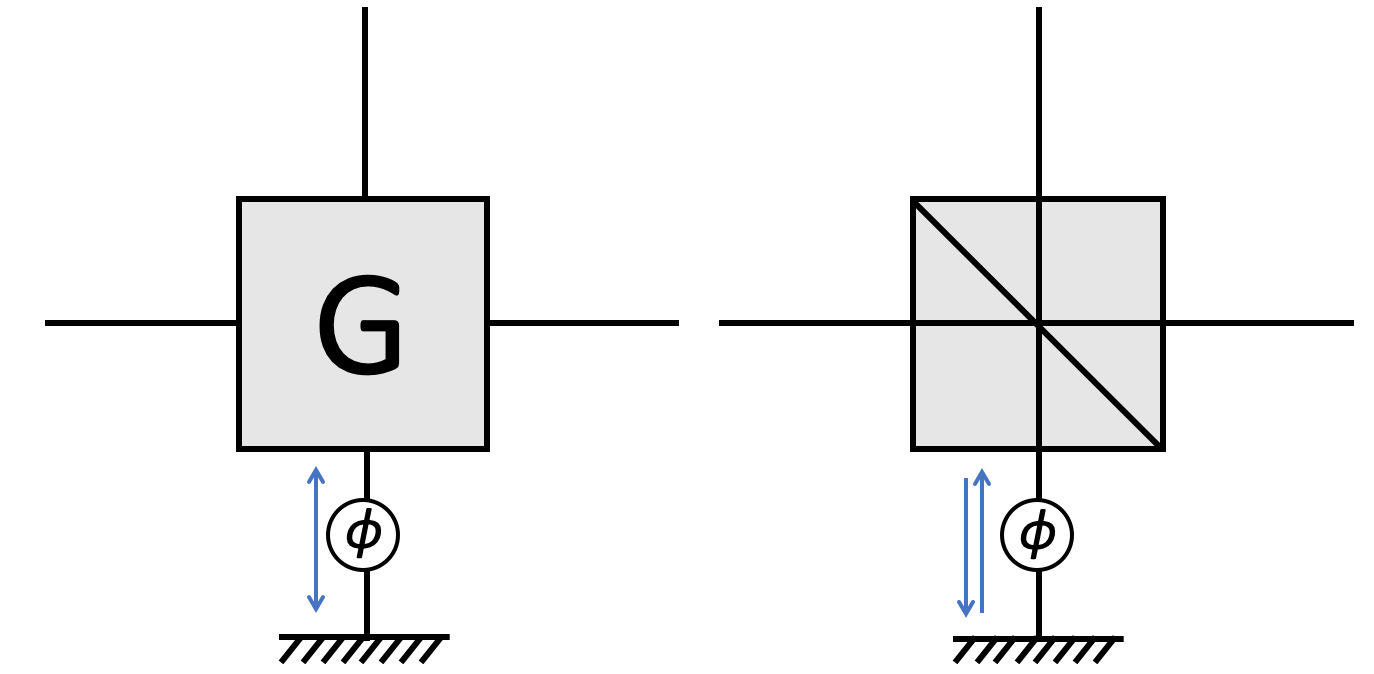}
\caption{(left) A simple optical cavity coupled to a Grover coin (G). The spatial mode at the lower port is converted to a cavity mode by sealing the port with a mirror. Light in this mode oscillates back and forth, depicted schematically with the blue arrow. \textit{As the light bounces within the cavity, some energy leaks into the output ports, interfering with the radiation in these ports.} As $\phi$ changes, the interference in the output ports changes, modulating the final scattering matrix. (right) The same topology but with a traditional beam-splitter instead of a Grover coin. \textit{Because the beam-splitter is directionally-biased, this configuration does not form an optical cavity.} All energy will leave the sealed port mode after a single round trip.
\label{fig:simple-seal}}
\end{figure}

Consider the configuration with one simple cavity affixed to a 4-port Grover coin, such as the sealed 4-port Grover coin pictured in Fig. \ref{fig:simple-seal}. Light entering the cavity strikes the mirror, accumulating phase $\pi$ from the mirror in addition to a round-trip phase $\phi$. Since the device has three open ports, the scattering matrix will be $3\times 3$ and its elements will vary $2\pi$-periodically with $\phi$. After each pass through the cavity, 3/4 of the energy of the optical state is transferred from the cavity mode uniformly into the spatial modes of the open ports.

When computing the scattering matrix, one can avoid making redundant calculations using the permutation symmetry of the Grover coin; after computing the output amplitudes for input at one port, the output state for input at a different port is the same but with the corresponding amplitudes permuted. In this example, the cavity mode amplitudes map for each round trip into a simple recursive form which can be converted to a geometric series. Ultimately, the reflected amplitude is of the form  $r = (e^{i\phi} - 2)^{-1}$ and the transmitted amplitude is $t = 1 + r$. One may readily verify these amplitudes are normalized: $|r|^2 + 2|t|^2 = 1$. For $\phi = 0$, the device behaves as a $3$-sided mirror. 

More interestingly, for $\phi = \pi$, the reflection probability reaches a minimum and the device behaves as a Grover $3$-port. It can be shown this property generalizes to Grover coins of arbitrary dimension $d$: a single seal cavity will create a device that interpolates between a $(d-1)$-port mirror and $(d-1)$-port Grover coin. Conversely, if one coherently couples an $n$-port Grover coin to a $d$-port Grover coin with a net 0 phase shift (modulo $2\pi$) in the cavity, a $(d+n-2)$-port Grover coin is formed. Applications of these facts will be explored elsewhere, and in the remainder of this article we focus on applications with low-dimensional devices.

\begin{figure}[ht]
\centering\includegraphics[width=0.45\textwidth]{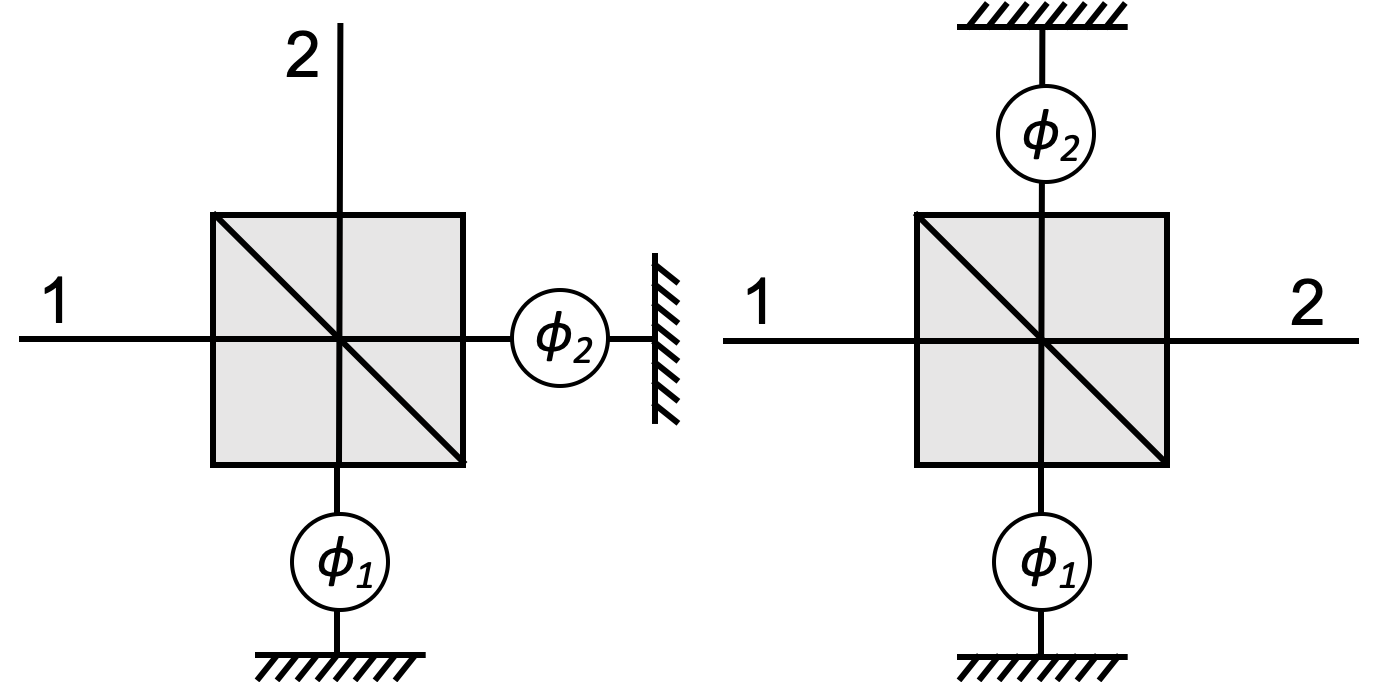}
\caption{(left) Traditional Michelson interferometer. (right) A cavity-coupled beam-splitter device. The scattering coefficients for both of these devices can be expressed as a function of the quantity $(\phi_1 + \phi_2)$ or $(\phi_1 - \phi_2)$.\label{fig:bs-devices}}
\end{figure}

The cavity formed in the previous example will not exist if a beam-splitter is used instead of a Grover coin. It will have a partially-unbiased scattering matrix: input at two ports will be directionally-unbiased while input at the third does not interact with the mirror and is consequently feed-forward. In fact, this device formed the essential building block of the original free-space realization of the Grover multiport \cite{PhysRevA.93.043845}\cite{Osawa:18}. 

Placing a pair of mirrors on adjacent sides of a beam-splitter will not form a cavity either but doing so will form another device of great practical use: the Michelson interferometer, shown in Fig. \ref{fig:bs-devices} (left). We can view this interferometer as a tunable, unbiased, two-port scattering device. Its reflection and transmission probabilities are given by
\begin{subequations}
\begin{align}
    R &= \cos^2\bigg(\frac{\phi_1 - \phi_2}{2}\bigg)\label{eq:m1}\\
    T &= 1 - R = \sin^2\bigg(\frac{\phi_1 - \phi_2}{2}\bigg)\label{eq:m2}
\end{align}
\end{subequations}
where $\phi_1$ is the phase in the first arm of the interferometer, and $\phi_2$ is that of the second.

If two mirrors are alternatively arranged to lie on opposite sides of the beam-splitter, as shown in Fig. \ref{fig:bs-devices} (right), an optical cavity is formed. The scattering matrix elements are found by coherently summing the probability amplitudes for all paths light can take between a given input port and a given output port. This device is also tunable, with the reflection and transmission probabilities given by 
\begin{subequations}
\begin{align}
    R &= \frac{1}{5 - 4\cos(\phi_1 + \phi_2)}\label{eq:bsR},\\
    T &= 1 - R.
\end{align}
\end{subequations}
In the case of using a conventional beam-splitter, the cavity and Michelson device probabilities may be expressed as functions of either $(\phi_1 + \phi_2)$ or $(\phi_1 - \phi_2)$. If $\phi_2$ is fixed at some value, the probabilities can be viewed as a $2\pi$-periodic curve parameterized by $\phi_1$, or vice versa. Altering $\phi_2$ then produces a new curve which is function of $\phi_1$. The points on any particular curve of fixed $\phi_2$ are prescribed for each value of $\phi_1$. {\color{black} Due to the functional dependence of the form $(\phi_1 \pm \phi_2)$}, when the fixed phase $\phi_2$ is allowed to vary, \textit{the entire transmission curve is merely translated along the horizontal axis without making changes to its structure}. Hence an entire degree of freedom is redundant. In addition to this, the cavity-based beam-splitter device cannot be tuned to obtain all reflection and transmission probabilities in the range $[0, 1]$; this can be seen since the minimum in Eq. (\ref{eq:bsR}) occurs when the denominator is maximized, which in turn occurs when the cosine term equals its extreme value of -1, so the minimum reflection probability obtainable is 1/9.

\section{Enhanced sensitivity Grover-Michelson interferometer}
\begin{figure}[ht]
\centering\includegraphics[width=0.4\textwidth]{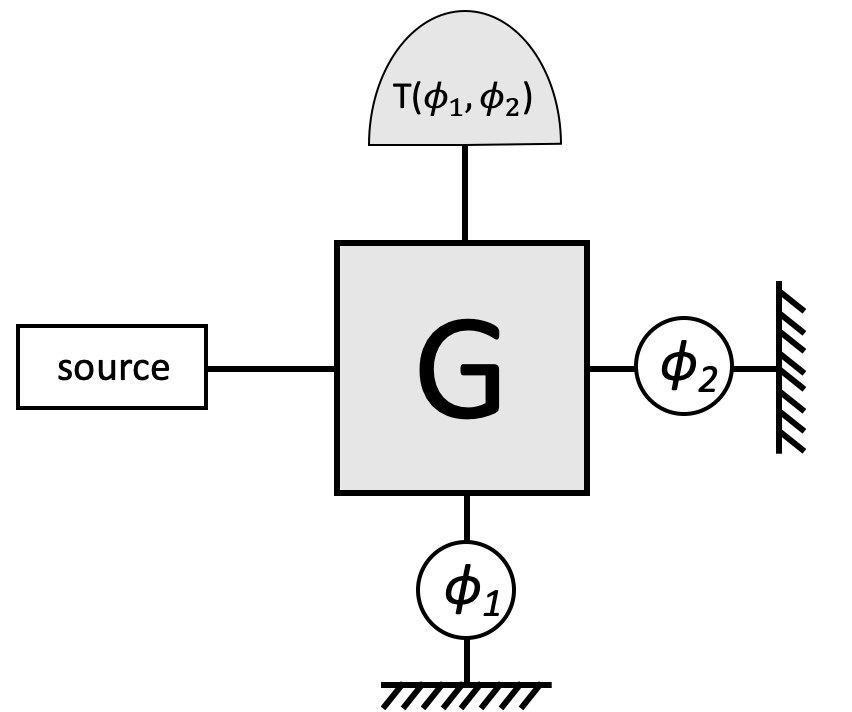}
\caption{Grover-Michelson interferometer, formed by replacing the beam-splitter in a conventional Michelson interferometer with a  Grover coin. By attaching a source and detector to the open ports, an enhanced phase sensing device is formed; the interference of cavity amplitudes leads to a nonlinear, continuously modified phase response, allowing the slope of the output probabilities with respect to a phase perturbation be made as steep or flat as desired.\label{fig:4p2s}}
\end{figure}

A novel two-port device can be created from the 4-port Grover coin by forming separate cavities at two ports. The device topologies in Fig. \ref{fig:bs-devices} coalesce into this ``Grover-Michelson" device when the beam-splitter is replaced by a Grover coin, as the latter device is permutation symmetric. The resulting configuration is shown schematically in Fig. \ref{fig:4p2s}. Light from a coherent source enters a port and the transmission probability is measured at the other open port. Some of the incident energy is coupled into two cavities, reflecting from a mirror and re-entering the Grover coin after accumulating a phase shift $\phi_j$ in cavity $j$.

Assume the cavities are coupled to ports 3 and 4. If one studies the round-trip propagation of $(a_3^\dagger + a_4^\dagger)$ and $(a_3^\dagger - a_4^\dagger)$ rather than $a_3^\dagger$ and $a_4^\dagger$ individually, the transformation from one round-trip to the next forms a recurrence relation which may be unrolled into a geometric series and then explicitly summed (see Appendix B). Thus these linear combinations somewhat emulate the role of coupled-cavity supermodes. In the appendix we show the output state corresponding to $|\psi_0\rangle = a_1^\dagger|0\rangle$ is given by
\begin{align}\label{eq:2seals}
|\psi_{\text{out}}\rangle = \bigg [
\bigg (\frac{C(\phi_1, \phi_2)^2}{2B(\phi_1, \phi_2) - 2} - \frac{B(\phi_1, \phi_2)}{2} - \frac12\bigg )a_1^\dagger& \notag\\ + \bigg (\frac{C(\phi_1, \phi_2)^2}{2B(\phi_1, \phi_2) - 2} - \frac{B(\phi_1, \phi_2)}{2} + \frac12\bigg )a_2^\dagger&\bigg]|0\rangle,
\end{align}
where 
\begin{subequations}
\begin{align}
    B(\phi_1, \phi_2) &\coloneqq \frac12 (e^{i\phi_1} + e^{i\phi_2}) \\
    C(\phi_1, \phi_2) &\coloneqq \frac12 (e^{i\phi_1}-e^{i\phi_2}).
\end{align}
\end{subequations}
In fact, we  show in the appendix that up to a $\pi$ phase shift, $B$ and $C$ are respectively the $r$ and $t$ for the standard Michelson interferometer. Therefore the use of a Grover coin instead of a beam-splitter results in a direct nonlinear transformation of the device's scattering coefficients.

\begin{figure}
\includegraphics[width=0.5\textwidth]{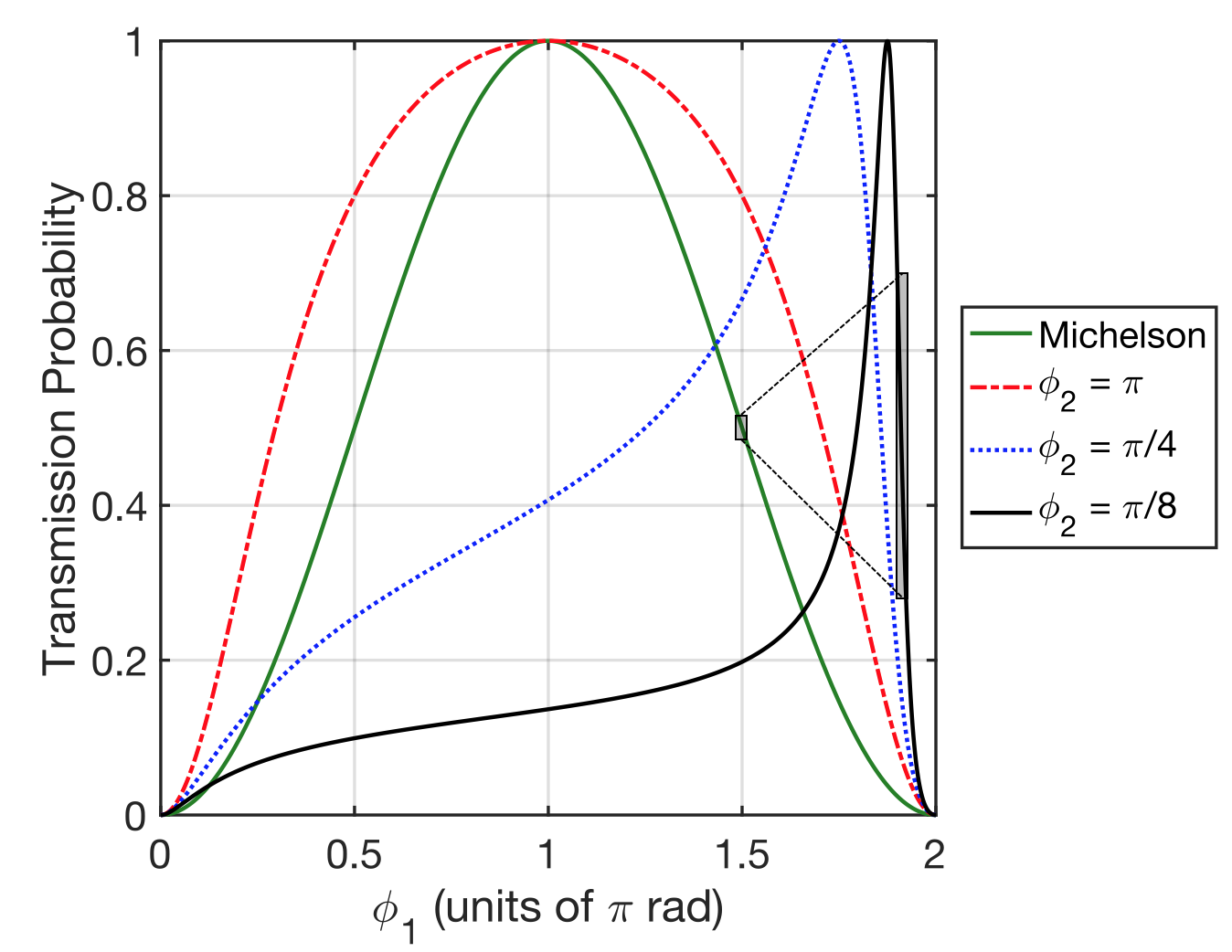}
\centering\caption{Transmission probability curve in one period $0 \leq \phi_1 \leq 2\pi$ for the Grover-Michelson interferometer at various values of $\phi_2$. At $\phi_2 = \pi$, $T$ is symmetric, but as $\phi_2$'s distance from $\pi$ is increased, the curve becomes increasingly skewed, leaving the periodic endpoints fixed. The homotopy originates from the fact that $R$ and $T$ are given by composition of nonlinear functions of $(\phi_1, \phi_2)$, which itself stems from the cavity-coupled nature of the device. Given suitable control of the phases $\phi_1$ and $\phi_2$, perturbations in the $\phi_1$ can be made to have a small or large response in $R$ and $T$, while in the standard Michelson the maximum slope is always 1. The gray boxes illustrate how this difference in modulation $\Delta T$ for the same phase variation $\Delta \phi$ can be made very large.
\label{fig:4p2s-curves}}
\end{figure}

{\color{black} Over the parameters $\phi_1$ and $\phi_2$, both the Michelson and Grover-Michelson interferometers span the line $R + T = 1$ with $R, T \geq 0$. This line forms the state space of classical scattering transformations derived from a tunable $U(2)$ device. Despite sharing this space entirely, the Grover-Michelson carries a significant advantage in the way its dependence on $\phi_1$ and $\phi_2$ covers the line $R + T = 1$.} To see this, consider how its reflection and transmission probability curves $R_{\phi_2}(\phi_1)$ and $T_{\phi_2}(\phi_1)$ vary with $\phi_2$, as in Fig. \ref{fig:4p2s-curves}. {\color{black} In the case of the regular Michelson interferometer, we recall from the previous section the scattering probabilities were functions of the quantity $(\phi_1 - \phi_2)$.} This meant that as one phase was varied, the probability curve was merely translated. 

However, in the Grover-Michelson device, the dependence on $\phi_1$ and $\phi_2$ cannot be expressed in this way. The geometric series summation of cavity amplitudes places phase dependence in the denominator in the output amplitudes, resulting in nonlinear behavior in $R$ and $T$. Examples of reflection probability curves are plotted in Fig. \ref{fig:4p2s-curves}. The deformation of each probability curve constrained on the domain $[0, 2\pi]$ is over a family of continuous paths with fixed endpoints, which is an instance of homotopy. At $\phi_2 = \pi$ the tuning curves are symmetric about $\phi_1 = \pi$. {\color{black} However, as $\phi_2$ increases in distance from $\pi$, the curves are increasingly skewed toward the periodic endpoint, increasing the maximum slope obtainable on a given curve.}

\begin{figure}
\includegraphics[width=0.5\textwidth]{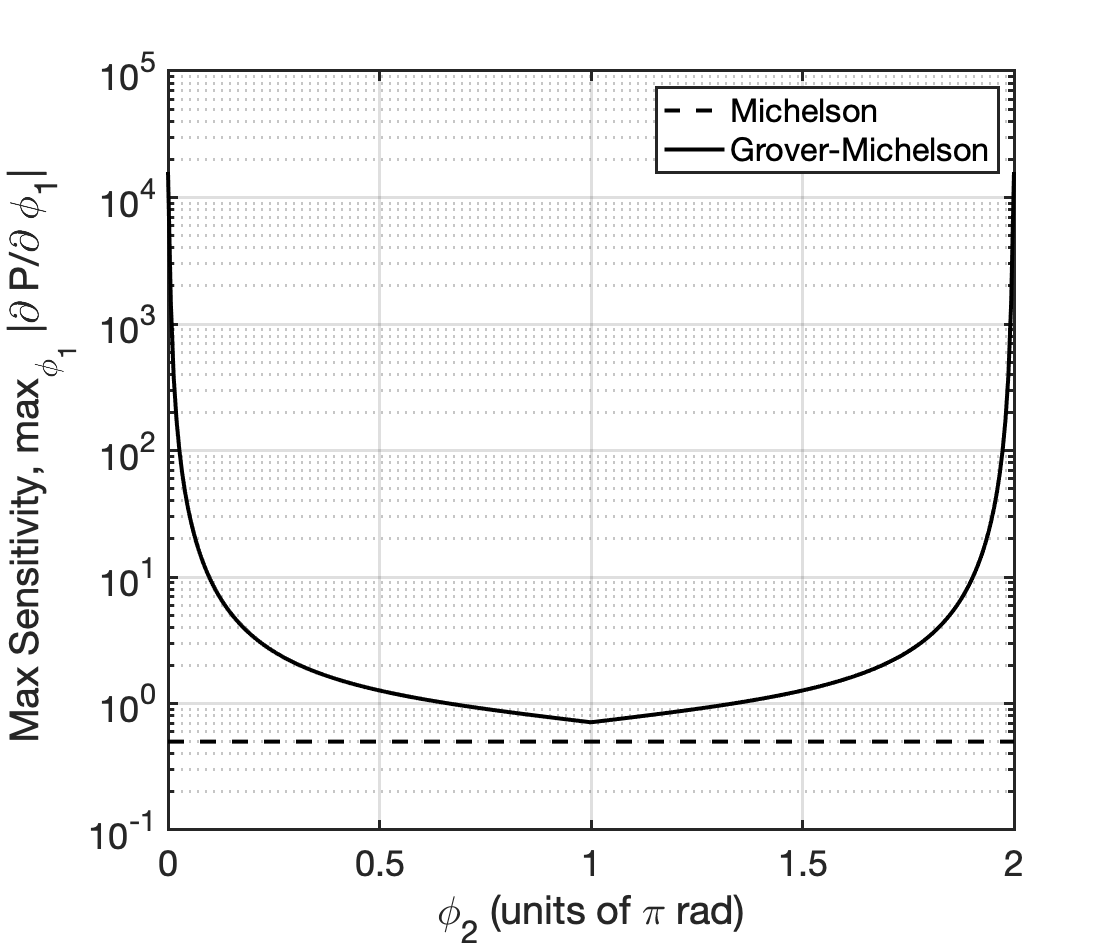}
\centering\caption{\label{fig:sens}Maximum sensitivity of the Michelson and Grover-Michelson interferometers versus $\phi_2$, shown from $10^{-5}$ to $2\pi - 10^{-5}$. The curves are periodic in $2\pi$. The sensitivity is given by the absolute value of the slope of either the reflection or transmission probability $P$, $|\partial P_(\phi_1, \phi_2)/\partial\phi_1|$. The maximum is taken over $\phi_1$. The Grover-Michelson dominates the Michelson for all values of $\phi_2$ and becomes arbitrarily large as $\phi_2$ approaches integral multiples of $2\pi$.}
\end{figure}

An advantage of the Grover-Michelson configuration over the traditional one is that the sensitivity of the output state can be made arbitrarily large or small by fixing $\phi_2$ at some value near an integer multiple of $2\pi$. Sensitivity here is quantified by the magnitude of the slope of the probability $R$ or $T$ vs. $\phi_1$ curve, such as $|\partial T/\partial \phi_1|$. A comparison of the maximum sensitivity as a function of the curve index $\phi_2$ is shown in Fig. \ref{fig:sens}. The phase sensitivity can be seen to grow arbitrarily large as $\phi_2$ approaches integral multiples of $2\pi$. 

From a metrological standpoint, the controllable sensitivity behavior can be used to substantially increase the resolution of the phase readout. Assuming the phase shift $\phi_2$ can be stably controlled, then this sharper response can be useful if the phase $\phi_1$ is perturbed about the sensitive region by a small, unknown amount. If the interferometer is calibrated with a controllable phase $\phi_1$ to bias the system about the point of 50:50 power splitting, say in the case $\phi_2 = \pi/8$ in Fig. \ref{fig:4p2s-curves}, then a small unknown variation in phase $\delta$ brings the effective cavity round-trip phase $\phi_1$ to $({\color{black}\phi_1} + \delta)$. Accordingly, the reflectance and transmittance will see a substantially greater degree of modulation in comparison to this same variation affecting a standard Michelson interferometer. This is illustrated with the shaded gray boxes in Fig. \ref{fig:4p2s-curves}; the boxes illustrate the same value of unknown phase disturbance leads to a substantially larger modulation in transmittance for the Grover-Michelson.

{\color{black} Using a phase-shifting element in the second arm to control $\phi_2$, the slope at a given bias point can be changed, allowing the system to be field-programmed to accommodate varying perturbation strengths, even if the strengths themselves are unknown. If the slope is large enough for a given perturbation, the output transmission will jump past the sensitive region and saturate into the flat-sloped, low-transmission region of the next period, at which readout of the phase becomes difficult to resolve accurately. This exemplifies the trade-off between sensitivity and dynamic range intrinsic to metrology. The larger the slope, the smaller $\Delta \phi$ must be in order to be extracted unambiguously. Assuming the same perturbation can be reapplied at will, one may repeatedly re-calibrate the bias point at different slopes and then reapply the perturbation here until the result is non-saturating. For large perturbations, this may mean operating in a low-curvature bias point to obtain a larger dynamic range. For instance, in Fig. \ref{fig:4p2s-curves}, biasing at the inflection point $\phi_1 = \pi, \phi_2 = \pi/8$ in the Grover-Michelson interferometer would be the preferred readout location in comparison to the $\phi_1 = \pi$ zero-sloped region of high curvature in the traditional device.

As a coherent device, conditions must be met to guarantee the device behaves as predicted by the assumption that amplitudes corresponding to indistinguishable paths interfere. {\color{black} As stated in Section II}, the source must be coherent with respect to the length scales of the Grover coin in use. This allows the physical Grover coin such as the one in Fig. \ref{fig:grover-real} to behave as such. The conditions pertaining to that physical realization are discussed in {\color{black} Ref.} \cite{PhysRevA.93.043845}. The coherence length of the source must also be larger than the length of a single cavity arm so that amplitudes exiting after a \textit{different} number of cavity round-trips can interfere. Otherwise, the result would be a statistical mixture of the output amplitudes corresponding to interference over an \textit{equal} number of round-trips only. Fortunately, the geometric series that determines the output ampltiudes converges quickly, so interference over only a few round-trips would likely be sufficient in practice.

Nonetheless, if the source coherence may be an issue, the condition may be improved by bringing the arms closer to the coin; however, if the arms become too close to the coin, they will couple to the internal cavity modes of the coin, thereby changing the coin to a new device of its own. Instead, one might ensure the coin acts instantaneously on each round-trip by using a detector with a slower response then characteristic lifetime of the coin's internal cavity. Then the time spent within the coin becomes negligible in relation to the time spent inside a cavity arm.

Other configurations have been known to impart a large, possibly nonlinear phase due to the interference of light in a cavity. For instance, a Fabry-Perot could provide a collinear cavity to be placed in the arm of a Michelson interferometer. If the back surface is fully reflective, a Gires-Tournois interferometer is formed \cite{GTI}\cite{Dingel:98}. While such a device can be designed to produce a sharp response to a phase perturbation, the curve it produces stems from fixed parameters such as surface reflectivity, thickness, and refractive index. This means the device can only produce one curve for each wavelength. One substantial advantage of the tunable response curves of the Grover-Michelson device is that any fixed wavelength can be used with the device, so long as the unbiased coin continues to act according to Eq. (\ref{eq:grover}). At the new wavelength, one obtains the same family of curves, by adjusting values of $\phi_2$.}

\section{Conclusion}
{\color{black} In this work we have shown that replacing the beam-splitter in a standard Michelson interferometer with a Grover coin generates a tunable-sensitivity device.} The new device behavior is the result of field interference over an infinite number of cavity round-trip paths, which generates a nonlinear phase dependence in the scattering matrix amplitudes. As a result, the transmittance can be made as sensitive or insensitive as desired by tuning an external phase and can operate at any fixed wavelength. This would provide an enhanced-resolution measurement of phase disturbances in any physical situation. Higher dimensional extensions of this work will be explored in the future.

\section*{Acknowledgments}
This research was supported by the Air Force Office of Scientific Research MURI award number FA9550-22-1-0312.

\section*{Appendix}
Here we derive the scattering amplitudes for the standard Michelson and Grover-Michelson interferometers. We work in the Heisenberg picture, in which the scattering matrix acts on the {\color{black} photon} creation operators. This is equivalent to the Schrödinger picture in which the probability amplitudes of each operator are transformed under the same mappings. Indeed, if a monochromatic but otherwise general optical scattering state is given by 
$$|\psi\rangle = \sum_{j}c_ja_j^\dagger |0\rangle = \sum_{j,k} c_ja^\dagger_k \delta_{jk}|0\rangle,$$
then applying the linear scattering transformation $A$ on $|\psi\rangle$ gives
$$A|\psi\rangle = \sum_{j, k, \ell} (A_{k j} c_j) a^\dagger_\ell \delta_{j \ell} |0\rangle = \sum_{j, k, \ell} c_j (A_{k \ell} a^\dagger_\ell) \delta_{j \ell}|0\rangle.$$

\subsection{Standard Michelson Interferometer}
We use the beam-splitter scattering matrix in Eq. (\ref{eq:bs}). Because the beam-splitter is biased there is no cavity summation. Thus the initial excitation is split into each arm, hits the mirror ($M$) and phase shifts ($\Phi)$ once, and then overlaps at the beam-splitter again before exiting. In accordance with this, we see
\begin{align*}
a_1^\dagger &\xrightarrow{B} \frac{1}{\sqrt{2}} (a_3^\dagger + a_4^\dagger) \\ &\xrightarrow{\Phi, M} -\frac{1}{\sqrt{2}} ( e^{i\phi_1} a_3^\dagger + e^{i\phi_2} a_4^\dagger ) \\ &\xrightarrow{B} -\frac12 \bigg ( ( e^{i\phi_1} (a_1^\dagger + a_2^\dagger) + e^{i\phi_2} (a_1^\dagger - a_2^\dagger) ) \bigg )\\ &= -\frac12 \bigg (a_1^\dagger (e^{i\phi_1} + e^{i\phi_2}) + a_2^\dagger(e^{i\phi_1} - e^{i\phi_2}) \bigg)
\end{align*}
Reading off the scattering amplitudes, 
\begin{align*}
r &= -\frac12 (e^{i\phi_1} + e^{i\phi_2}),\\
t &= -\frac12 (e^{i\phi_1} - e^{i\phi_2}).
\end{align*}
The square-modulus of these leads to the scattering probabilities of eqs. (\ref{eq:m1}) and (\ref{eq:m2}). A similar calculation starting with $a_2^\dagger$ results in the same output amplitudes.
\subsection{Grover-Michelson Interferometer}
In a Grover coin $G$, we will seal port 3 with a phase shift $\phi_1$ and mirror and do the same for port 4 with a phase shift $\phi_2$ and mirror. Collectively the linear phase transformations these devices enact during each round trip will be denoted $\Phi$ and $M$. To simplify the calculation, we will introduce some new variables and first show how a single round-trip affects $(a_3^\dagger + a_4^\dagger)$ and $(a_3^\dagger - a_4^\dagger)$. We will find that these linear combinations of cavity modes are mapped recursively into themselves, somewhat emulating the role of coupled-cavity supermodes.

To that end, define the following:
\begin{align*}
    A &\coloneqq \frac12(-a_1^\dagger + a_2^\dagger), \\ 
    B &\coloneqq \frac12 (e^{i\phi_1} + e^{i\phi_2}), \\
    C &\coloneqq \frac12 (e^{i\phi_1}-e^{i\phi_2}).
\end{align*}
Next, for an excitation $(a_3^\dagger + a_4^\dagger)$ making a single round-trip,
\begin{subequations}
\begin{align}
(a_3^\dagger + a_4^\dagger) &\xrightarrow{M, \Phi} -e^{i\phi_1}a_3^\dagger -e^{i\phi_2}a_4^\dagger \\&\xrightarrow{G} -\frac12 (e^{i\phi_1}(a_1^\dagger + a_2^\dagger - a_3^\dagger + a_4^\dagger) \notag\\ & \ +  e^{i\phi_2}(a_1^\dagger + a_2^\dagger + a_3^\dagger - a_4^\dagger)) \\&= -\frac12 (e^{i\phi_1} + e^{i\phi_2})(a_1^\dagger+a_2^\dagger) \notag\\& \ + \frac12 (e^{i\phi_1}-e^{i\phi_2})(a_3^\dagger - a_4^\dagger) \\ &= -B(a_1^\dagger+a_2^\dagger) + C(a_3^\dagger - a_4^\dagger)\label{eq:1}
\end{align}
\end{subequations}
and similarly
\begin{subequations}
\begin{align}
(a_3^\dagger - a_4^\dagger) &\xrightarrow{M, \Phi} - e^{i\phi_1}a_3^\dagger + e^{i\phi_2}a_4^\dagger \\ &\xrightarrow{G} -\frac12(e^{i\phi_1}(a_1^\dagger + a_2^\dagger - a_3^\dagger + a_4^\dagger) \notag\\&-e^{i\phi_2}(a_1^\dagger + a_2^\dagger + a_3^\dagger - a_4^\dagger)) \\ &= -\frac12(e^{i\phi_1} - e^{i\phi_2})(a_1^\dagger + a_2^\dagger) \notag\\ & \ + \frac12 (e^{i\phi_1} + e^{i\phi_2})(a_3^\dagger - a_4^\dagger) \\ &= -C(a_1^\dagger + a_2^\dagger) + B(a_3^\dagger - a_4^\dagger).\label{eq:2}
\end{align}
\end{subequations}
We see $(a_3^\dagger - a_4^\dagger)$ maps directly into itself after each round trip. This recursion can be explicitly unrolled into a geometric series and summed like so
\begin{subequations}
\begin{align}
(a_3^\dagger - a_4^\dagger) &\xrightarrow{N} -C (a_1^\dagger + a_2^\dagger) \sum_{n=0}^N B^{n} + B^{N+1}(a_3^\dagger - a_4^\dagger) \\
&\xrightarrow{N\rightarrow\infty} - C(a_1^\dagger + a_2^\dagger)\sum_{n=0}^\infty B^n  \\
&= \bigg(\frac{C}{B - 1} \bigg )(a_1^\dagger + a_2^\dagger).\label{eq:3}
\end{align} 
\end{subequations}
Now we use the above to derive the S-matrix. The Grover coin maps a photon incident on the first port to
\begin{align*}
    a_1|0\rangle &\rightarrow \frac12 (-a_1^\dagger + a_2^\dagger + a_3^\dagger + a_4^\dagger)|0\rangle = (A + \frac12 (a_3^\dagger + a_4^\dagger))|0\rangle.
\end{align*} 
Combining this with the above formulas, we see
\begin{align*}
&\xrightarrow{(\ref{eq:1})} \bigg[A - \frac{B}{2}(a_1^\dagger + a_2^\dagger) + \frac{C}{2}(a_3^\dagger - a_4^\dagger)\bigg ]|0\rangle \\ &\xrightarrow{(\ref{eq:3})}\bigg[A - \frac{B}{2}(a_1^\dagger + a_2^\dagger) + \frac{C}{2}\bigg(\frac{C}{B - 1}\bigg )(a_1^\dagger + a_2^\dagger)\bigg]|0\rangle.
\end{align*}
Grouping by operator yields the output state (\ref{eq:2seals})
\begin{align}
|\psi_{\text{out}}\rangle =\bigg [
\bigg (\frac{C^2}{2B - 2} - \frac{B}{2} - \frac12\bigg )a_1^\dagger &\notag\\ + \bigg (\frac{C^2}{2B - 2} - \frac{B}{2} + \frac12\bigg )a_2^\dagger&\bigg]|0\rangle.
\end{align}
Up to a $\pi$ phase shift, the $B$ and $C$ are respectively $r$ and $t$ for the traditional Michelson. Hence the above calculation illustrates that the Grover coin nonlinearly maps the scattering parameters of the standard Michelson interferometer. Because the Grover coin is permutation symmetric there is no need to consider the initial state $a_2^\dagger |0\rangle$. Relabeling ports $1 \longleftrightarrow 2$ results in the same permutation of the output amplitudes so that $r$ and $t$ are the same for input on either port.
\bibliography{refs}

\end{document}